\documentclass[twocolumn, prl]{revtex4}

\usepackage{graphicx}
\usepackage{times}
\usepackage{dcolumn}
\usepackage{bm}

\begin{document}

\title{ Disordered and Ordered States in a Frustrated Anisotropic 
Heisenberg Hamiltonian}

\author {S. Moukouri$^{1,2}$ and J.V. Alvarez$^1$}

\affiliation{ $^1$Department of Physics and  $^2$Michigan Center for 
          Theoretical Physics \\
         University of Michigan 2477 Randall Laboratory, Ann Arbor MI 48109}

\begin{abstract}
We use a recently proposed perturbative numerical renormalization group
algorithm to investigate ground-state properties of a frustrated  
three dimensional Heisenberg model on an anisotropic lattice. We analyze the 
ground state energy, the finite size spin gap and the static magnetic
structure factor. We find in two dimensions a frustration-induced gapless 
spin liquid state which separates two magnetically ordered phases. 
In the spin liquid state, the  magnetic structure factor  shows evidence 
that this state is made of nearly disconnected chains. This
spin liquid state is unstable against unfrustrated interplane couplings.  
\end{abstract}

\maketitle
Low dimensional quantum magnets are currently the object of an
important interest \cite{review}. This stems from the rich physics
which is displayed by these sytems due to their reduced dimension and
competing interactions which often push the transition to ordered
states to very low temperatures or even preclude the onset of 
magnetism at all. Geometric frustration is one of the effects
which are believed to lead to possible non-classical states. A
non-classical ground state exists in a pure one-dimensional quantum
antiferromagnet. The ground state is disordered and the low
energy excitations are spinons which carry fractional quantum numbers.
A fundamental question is whether spinons can survive in higher
dimension.

The relevance of these questions has been substantiated in a 
recent neutron scattering experiment \cite{coldea} reported in 
the quantum magnet Cs$_2$CuCl$_4$.
This compound is a quasi-two dimensional spin
one-half Heisenberg antiferromagnet. It is made of anisotropic
triangular planes (with $J'=J" \approx 0.34 J$ in Eq(\ref{heisenberg})) which
are weakly-coupled by an exchange which is roughly $J_z=0.045J$. An 
incommensurate N\'eel state is stable below $T_N=0.62 K$ at 
zero magnetic field. In
the two dimensional regime above $T_N$, the dynamic correlation displays
a highly dispersive continuum of excited states which is a signature of spinons.
This finding together with the fact that $J'$ is in the same order as $J$
led to the conclusion that above $T_N$, Cs$_2$CuCl$_4$ is a 2D spin liquid
with fractional excitations.

 In order to study these effects of frustration and dimensional
crossover, we will consider the following anisotropic 3D 
Heisenberg Hamiltonian:

\begin{eqnarray}
 \nonumber H=J \sum_{i,l,k}{\bf S}_{i,l,k}{\bf S}_{i+1,l,k}+J'
\sum_{i,l}{\bf S}_{i,l,k}{\bf S}_{i,l+1,k}+ \\
J" \sum_{i,l,k}{\bf S}_{i,l,k}{\bf S}_{i+1,l+1,k}+J_z \sum_{i,l,k}{\bf S}_{i,l,k}{\bf S}_{i,l,k+1}
\label{heisenberg},
\end{eqnarray}

\noindent where the $i$ index represents sites, the $l$ index chains and 
the $k$ index planes, the exchange couplings are such that $J_z \ll J=1$
and $J',J" \alt 1$. When $J' \neq 0$ and $J" \neq 0$  the system is
frustrated. In this case the Quantum Monte Carlo method, which is so far
the most reliable method of investigation of spin systems, is plagued by
the sign problem. Alternative approaches \cite{series,RPA}
have investigated the possibility of a spin liquid state in the 2D 
regime ($J_z=0$) of the Hamiltonian(\ref{heisenberg}).
 In Ref. \cite{series}, this 2D model was studied using a combination 
of Ising and dimer series expansions. Considering the extrapolation 
of their results to the limit of  weakly coupled chains,  
the authors  was not able to conclude clearly due to the limitation of their
technique. They argued that either a spiral ordered or a nearly critical 
disordered phase are possible in this regime. The dynamical susceptibility 
of this model was computed within the Random Phase Approximation 
(RPA) \cite{RPA} using  an essentially exact expression for the 
1D chain dynamical susceptibility. For the 2D system 
incommensurate order with exponentially small characteristic 
wave vector is predicted. But this prediction was inspired by the 
experimental observation. It is impossible to tell from this study
if this is an intrinsic behavior of the Hamiltonian(\ref{heisenberg}).


We wish to present in this letter an ab-initio computation of the
ground state static properties of the Hamiltonian(\ref{heisenberg}).
For this purpose, we use a recently proposed perturbative density matrix 
renormalization group (DMRG) approach 
\cite{moukouri-TSDMRG, moukouri-TSDMRG2, moukouri-KB,alvarez-ED}.
 This perturbative DMRG method which has been so far used for only 
anisotropic 2D systems  is extended here to anisotropic 3D systems. We
will explore the following issues: (i) what is the nature of the
ground state in the 2D regime of the Hamiltonian (\ref{heisenberg})?
(ii) Is a spin liquid state favored when $J'=J"=0.3$ as found 
experimentally? If so, what is the nature of this spin liquid
state? (iii) Does $J_z$ restore a magnetic state starting from this
eventual spin liquid state? If so, is this order incommensurate?

 The perturbative DMRG which will be used in this study is a particular 
case of a more general matrix perturbation method based on Kato-Bloch 
expansion \cite{kato,bloch} which was recently 
introduced by one of us \cite{moukouri-KB}. 
In the first step, the usual 1D DMRG method \cite{white}
 is applied to find a set of low 
lying eigenvalues $\epsilon_n$ and eigenfunctions $|\phi_n \rangle$ of a 
single chain. One can note that during this step, if one wishes to study 
not too large lattices, it is preferable to use the exact diagonalization 
method instead of the DMRG. The advantage of the exact diagonalization method 
is that by selecting both the total spin $S$ and the momentum in addition to 
the $S_z$ component of $S$ as done in DMRG, it will lead to a better 
estimation of the low energy Hamiltonian of a single chain.

In the second step, the  2D 
Hamiltonian (Eq.(\ref{heisenberg}) with $J_z=0$ )is then projected
onto the basis constructed from the tensor product of the $|\phi_n \rangle$'s.
This projection yields an effective one-dimensional Hamiltonian for a
single plane (we drop temporaly the plane index $k$),

\begin{figure}
\includegraphics[width=3. in]{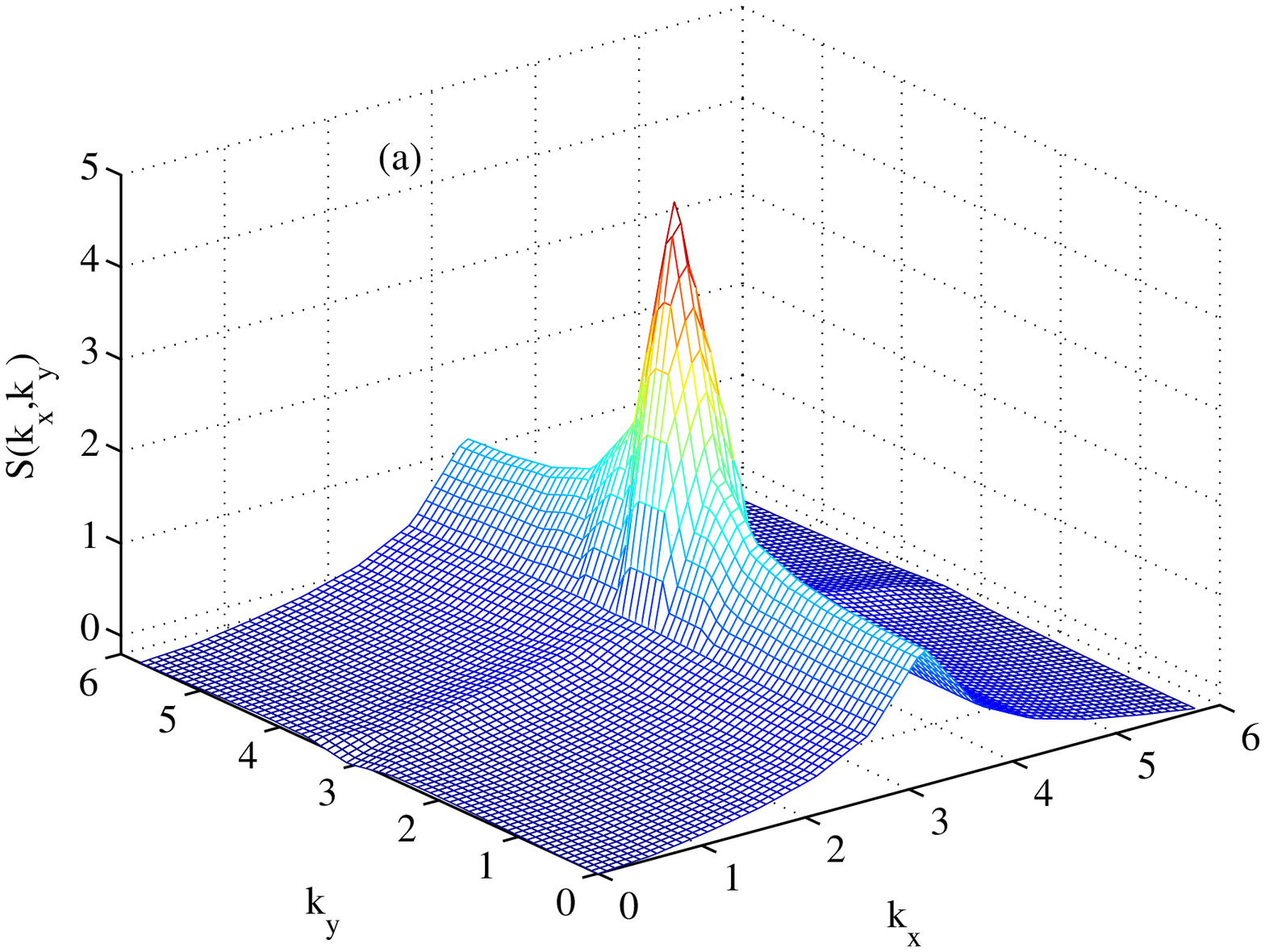}
\includegraphics[width=3. in]{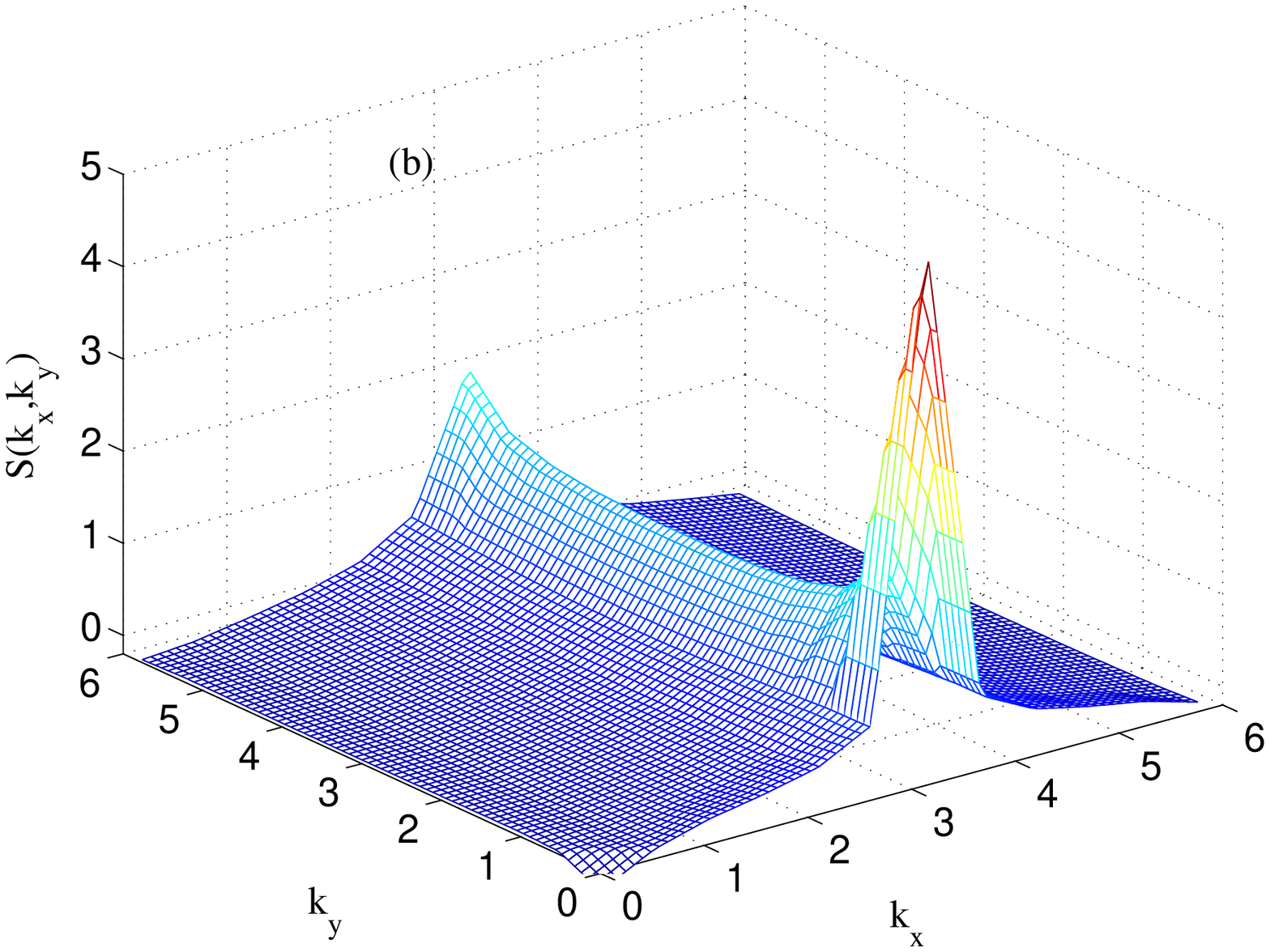}
\caption{ Static spin structure factor $S(k_x,k_y)$ for $L_x \times L_y = 24 \times 25$, $J'=0.3$, $J"=0$ (upper graph) and  $J'=0.3$, $J"=0.6$ (lower graph)}
\label{triang1}
\end{figure}

\begin{eqnarray}
\nonumber \tilde{H}_2 \approx \sum_{[n]} E_{[n]} 
|\Phi_{[n]}\rangle \langle \Phi_{ [n]}| +
 J' \sum_{l} {\bf \tilde{S}}_{l} {\bf \tilde{S}}_{l+1}+\\
J" \sum_{l} {\bf \tilde{S}}_{l}{\bf \tilde{S}}_{l+1}
\label{efhamil}
\end{eqnarray}

\noindent where  $E_{ [n]}$ is the sum of eigenvalues of the 
different chains, $E_{ [n]}=\sum{\epsilon_{n_l}}$; 
$|\Phi_{[n]}\rangle$ is the corresponding eigenstate, 
$|\Phi_{[n]}\rangle =  |\phi_{n_1}\rangle  |\phi_{n_2}\rangle ...  
|\phi_{n_L} \rangle$; the composite chain-spin operators on the 
chain $l$ are ${\bf \tilde{S}}_{l}=({\bf \tilde{S}}_{1l},
 {\bf \tilde{S}}_{2l}, ...{\bf \tilde{S}}_{Ll})$, where the running
index labels sites in a chain $l$. Note that the 
products of spin-chain operators in Eq.~\ref{efhamil} are different
for the terms involving $J'$ and $J"$. The term with $J'$ couples spins
with same site index on neighboring chains, when the one with  $J"$ 
couples spins with
site index $i$ on chain $l$ and $i-1$ on chain $l+1$ are coupled.   
 These renormalized matrix elements on the single chain basis are  

\begin{eqnarray}
 {\bf \tilde{S}}_{i,l}^{n_l,m_l}=\langle \phi_{n_l}|
{\bf S}_{i,l}|\phi_{m_l}\rangle
\end{eqnarray}

 The effective Hamiltonian (\ref{efhamil}) is one-dimensional and it 
is also  studied by the DMRG method. The only difference with a normal 
1D situation is that the local operators are now $m \times m$ matrices,
where $m$ is the number of states kept to describe the single chain.

In the third step, the same procedure is repeated in order to go from 2D 
to 3D once $J_z$ is set on. One then obtains the following effective 1D 
Hamiltonian for the 3D system:

\begin{eqnarray}
 \tilde{H}_3 \approx \sum_{[p]} E_{[p]}' 
|\Psi_{[p]}\rangle \langle \Psi_{ [p]}| +
 J_z \sum_{k} {\bf \tilde{S'}}_{k} {\bf \tilde{S'}}_{k+1}
\label{efhamil2}
\end{eqnarray}

\noindent where $E_{ [p]}'$ are the sum of eigenvalues of the decoupled
planes, $|\Psi_{[p]}\rangle$ are the corresponding eigenstate, and the
composite plane-spin operators on a plane $k$ are 
 ${\bf \tilde{S'}}_{k}=({\bf \tilde{S}}_{1k},{\bf \tilde{S}}_{2k}, ...
{\bf \tilde{S}}_{Pk})$, the running index labels chains in a plane $k$.
The generation of the effective Hamiltonian (\ref{efhamil2})
is identical to that of the effective Hamiltonian (\ref{efhamil}) described
in Refs.\cite{moukouri-TSDMRG, moukouri-TSDMRG2}. We target the spin
sectors $S_z=0$, $\pm1$, and $\pm 2$ to generate a low energy Hamiltonian
which describes the 2D systems.
In Ref\cite{moukouri-TSDMRG2, alvarez-ED}, this perturbative DMRG was 
tested against the stochastic series expansion quantum Monte Carlo (QMC) 
and exact diagonalization methods respectively. The agreement was very
good for small transverse couplings and not too large lattices. These
comparisons showed that the method is well controlled and its accuracy
can systematically be improved by increasing the number of states kept
during each step.

We first set $J_z=0$ and study the ground state properties of a
single plane. We studied lattice sizes ranging from $8 \times 9$ to 
$32 \times 33$ for $J'=0.15$ and $J"$ varying from $0$ to $0.3$; 
then $J'=0.3$ and $J"$ going from $0$ to $0.6$. We tipically keep
$m_x=128$ states in the single chain calculations; i.e. the whole chain
is described by $4 \times 128^2$ states.  Among these states we keep
only up to $m_y=64$ states in the second step; i.e. a plane is
described by $64^3$ states. The truncation errors are smaller
than $10^{-8}$. The reason for this abnormally small truncation 
error, which is related to the use of three blocks instead of four,
 has been discussed in Ref.\cite{moukouri-TSDMRG}.

 We found a similar qualitative behavior for these two sets of parameters. 
When $J" < J'$, the correlation in the transverse direction are AFM; 
the static structure factor $S(k_x,k_y)$ shown in the upper graph
of Fig.~\ref{triang1} 
for a $24 \times 25$ lattice, displays a maximum at $(k_x,k_y)=(\pi,\pi)$ 
(Note that the triangular lattice is equivalent to a square lattice with 
a coupling along one diagonal only. We thus adopt a square lattice notation 
throughout this study). We can thus conclude that in this regime, a 
N\'eel state is stable.  Now if $J" > J'$, the maximum in $S(k_x,k_y)$ 
shifts to $(\pi,0)$ as seen in the lower graph of  Fig.~\ref{triang1}. This is a signature of 
the collinear phase as in  Ref.\cite{moukouri-TSDMRG2}. This behavior was 
expected and has been found in other frustrated models: if one of the competing
exchange parameters is dominant, the corresponding ordered state
( the order which might prevail in the absence of the other competing
exchange) is favored \cite{chandra,dagotto}. 
A disordered state can only be expected in the region
of parameter space where the competing exchanges are close.

\begin{figure}
\includegraphics[width=3. in]{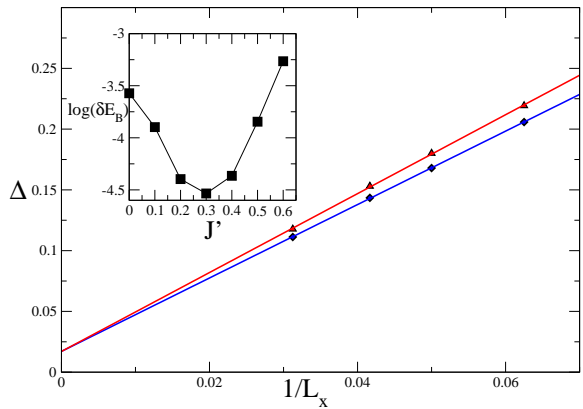}
\caption{Finite size spin gap of 2D lattices (triangles) versus that of
a single chain (diamonds). In the inset, error 
in the binding energy $\delta E_B=|E_B-E_B^{ED}|$ (in log scale) as a function of $J"$ for $J'=0.3$.}
\label{gaps}
\end{figure}

 We now concentrate on to the regime  $J"=J'$. The possibility of a 
disordered critical state was raised by the series expansion study of 
Ref\cite{series}. The expansion becomes inacurate for small values 
of $J'/J$, these authors could not reach this critical regime.
 Our data show the existence of such a disordered state  for
$J'=J"=0.15$, $0.3$, and $0.45$. Fig.~\ref{gaps} shows for $J"=J'=0.3$ 
the finite size gap $\Delta_{2D}$ of the 2D system and that of a single
chain $\Delta_{1D}$.  $\Delta_{2D}$ is found to be always
smaller than $\Delta_{1D}$. Since $\Delta_{1D} \rightarrow 0$ when
$L \rightarrow \infty$ and $m \rightarrow \infty$, we can thus 
conclude that the 2D system is also
gapless in the thermodynamic limit. The 1D version of this model, 
the two-leg zigzag ladder \cite{white-affleck}, has an 
exponentially small gap ($\Delta_{1D} \propto \exp(-\alpha J"/J)$ and $\alpha$ is a constant) hard to extrapolate from a finite size analysis.
That small gap coexists with 
ferromagnetic correlations between adjacent spins 
on the weakest bonds of the dimerized states. None of these 
concurrent signatures of the 1D frustrated system 
are observed in the 2D model where the ferromagnetic 
correlations favor a collinear state.

\begin{figure}
\includegraphics[width=3. in]{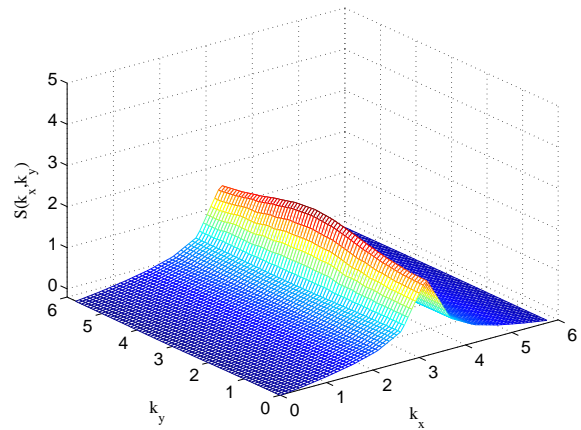}
\caption{$S(k_x,k_y)$ for $L_x \times L_y = 24 \times 25$, $J'=0.3$ and
 $J"=0.3$.}
\label{triang2}
\end{figure}

For $J'=J"$, $S(k_x,k_y)$ shown in Fig.~\ref{triang2} is structureless in the 
$k_y$ direction and it retains its maximum at $\pi$ in the $k_x$ direction.
This indicates that the chains are disconnected. The small bump
which can be observed in $S(k_x,k_y)$ near $(\pi,\pi)$ is a consequence
of very small short-range transverse AFM correlations. Thus the spin
liquid state is mostly 1D. Our conclusion of a disconnected chains
ground state is also supported by the transverse local bond strength
$|\langle {\bf S}_{il}{\bf S}_{il+1} \rangle |$ and the binding energy
of the chains $E_B=E_{1D}-E_{2D}/(L+1)$ 
(not shown), where $E_{1D}$ is the ground
state energy of a chain of length $L$ and $E_{2D}$ the ground state
energy of a $L \times L+1$ lattice. 
$|\langle {\bf S}_{il}{\bf S}_{il+1} \rangle |$  decays from $0.0818$
in the N\'eel state ($J'=0.3$, $J"=0$) and $0.0766$ in the collinear
state ($J'=0.3$, $J"=0.6$) to $0.0132$ in the disordered state 
($J'=0.3$, $J"=0.3$).

Indeed, one expects the perturbative DMRG method to be valid
only at small couplings.
 One may thus question its validity for
relatively large couplings such as $J'=J"=0.3$, and $0.45$. 
In Ref\cite{alvarez-ED} we have shown that, when larger 
values of the transverse coupling are used, the two-step DMRG is indeed 
less accurate when a genuine two-dimensional state emerges. But at the 
maximally frustrated point
(which is at $J'=J"$) the method remains surprisingly
accurate even for intermediate values of $J'$ and $J"$. 
This is seen in the inset of  Fig.\ref{gaps} which shows that $ \delta E_B$, the
difference in binding energy between the perturbative DMRG and exact
diagonalization, is minimum at $J'=J"$.
This is because
at the maximally frustrated point, the interchain correlations are
very small. Thus consistent with the observation made on $S(k_x,k_y)$, 
in the ground state, the eigenstates of an isolated chains are a good
approximation for the 2D system. 
 
\begin{figure}
\includegraphics[width=3. in]{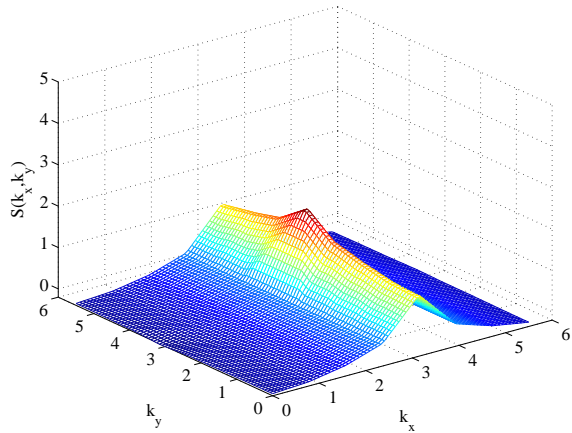}
\caption{$S(k_x,k_y)$ for $L_x \times L_y = 16 \times 16$, $J'=0.3$,
 $J"=0.3$ and $J_z=0.15$.}
\label{triang4}
\end{figure}

We now study the stability of the 2D spin liquid state against interplane
perturbations. 
 The 3D results are less accurate than the 2D ones.
The maximum truncation error was about 
$10^{-3}$ during the generations of the effective 2D Hamiltonian to be
used as the starting point in the 3D computations. This relatively
large truncation error is due to the fact that five spin sectors 
are targeted. We thus were restricted to smaller 2D lattices. 
 The largest lattice we studied was $16 \times 16 \times 17$. We set 
  $J'=J"=0.3$ and we vary $J_z$ from $0.05$ to $0.15$. 
Fig.~\ref{triang4} displays $S(k_x,k_y)$ for $J_z=0.2$ in the middle
plane, i.e. the $9^{th}$ plane in a $16 \times 16 \times 17$ lattice. 
One can observe the 
emergence of a small peak at $(\pi,\pi)$ which is due to
the interplane coupling. We find that as one may expect, the height of 
this peak increases with increasing $J_z$. Since the coupling in the
third direction is not frustrated, magnetic energy can be gained by
coupling the planes. As a consequence, the spin liquid state which
results from the inability of the chains within the planes to couple
effectively is destroyed.

To summarize,  we have studied the ground state properties
 of a 3D anisotropic  Heisenberg model. We can now address
the issues mentioned in the introduction which were motivated by
the experimental results of Ref.\cite{coldea}.
(i) In the 2D regime of this model, the magnetic structure factor, 
the binding energy  display the properties of a spin liquid state.
This spin liquid state separate two regimes with long-range order.
(ii)  This spin liquid state is gapless and is made of nearly 
independent chains. It is remarkable that it survives
even when the interchain exchange are of the same magnitude as the
intrachain exchange. Since it has been previously reported in the
crossed chain model \cite{singh} and on a
square lattice, it seems to be generic of models of chains
coupled with a frustrating interaction. (iii) Turning
on the interplane coupling restores magnetism.  But we do not
find any tendency to incommensurate ordering. The discrepancy on
this point between our simulation and the experiment \cite{coldea} may
be due to the fact that in the compound Cs$_2$CuCl$_4$, consecutives
triangular layers are sligthly rotated from each other. This
effect can be taken into account by adding a  Dzyaloshinskii-Moriya interaction
into the Hamiltonian(\ref{heisenberg}). This term could be the source of
the incommensurate ordering.

\begin{acknowledgments}
We thank Jim Allen for a critical reading of our manuscript.
\end{acknowledgments}


\begin{thebibliography}{99}
\bibitem{review}M.F. Collins and O.A. Petrenko, Can. J. Phys. {\bf 75},
                605 (1997).
\bibitem{coldea} R. Coldea {\it et. al.}, Phys. Rev. Lett. {\bf 86},
 1335 (2001); Phys. Rev. {\bf B 68}, 134424 (2003).
\bibitem{series} Z. Weihong, Ross H. McKenzie, R. R. P. Singh
            Phys. Rev. {\bf B 59}, 14367 (1999).
\bibitem{RPA} M. Bocquet, F.H.L. Essler, A. M. Tsvelik, A. O. Gogolin
            Phys. Rev. {\bf B 64}, 094425 (2001).
\bibitem{moukouri-TSDMRG} S. Moukouri and L.G. Caron, Phys. Rev. {\bf B 67},
                 092405 (2003).
\bibitem{moukouri-TSDMRG2} S. Moukouri cond-mat/0305608 (unpublished).
\bibitem{moukouri-KB} S. Moukouri physics/0312011 (to appear in Phys. Lett. A).
\bibitem{alvarez-ED} J.V. Alvarez and S. Moukouri cond-mat/0402530 (unpublished).
\bibitem{kato}T. Kato, Prog. Teor. Phys. {\bf 4}, 514 (1949); {\bf 5}, 95
              (1950).
\bibitem{bloch} C. Bloch, Nucl. Phys. {\bf 6}, 329 (1958).
\bibitem{white} S.R. White, Phys. Rev. Lett. {\bf 69}, 2863 (1992). Phys.
              Rev. {\bf B 48}, 10 345 (1993).
\bibitem{dagotto} E. Dagotto and A. Moreo, Phys. Rev. Lett. {\bf 63},
             2148 (1989).
\bibitem{chandra}P. Chandra, P. Coleman and A.I. Larkin, Phys. Rev. Lett.
                 {\bf 64}, 88 (1990).
\bibitem{white-affleck} S.R. White and I. Affleck, Phys. Rev. {\bf B 54},
9862 (1996).
\bibitem{singh} O.A. Starykh, R.R.P. Singh and G.C. Levine, Phys. Rev. Lett.
             {\bf 88}, 167203 (2002).


\end{thebibliography}
\end{document}